\documentstyle[times,graphics,epsfig,amsmath,amssymb,natbib]{mn2e}

\newcommand{\be}{\begin{equation}}
\newcommand{\e}{\end{equation}}
\newcommand{\bear}{\begin{eqnarray}}
\newcommand{\ear}{\end{eqnarray}}
\newcommand{\nline}{\nonumber \\}

\newcommand{\de}{{\rm d}}

\def\apj{ApJ}

\def\mnras{MNRAS}

\def\u{{\vec U}}
\def\th{\vec{\theta}}

\begin{document}
  \title[Simulating matched filter search for ionized
    bubbles]{Simulating the impact of HI fluctuations on  matched
    filter search for ionized bubbles in redshifted 21 cm maps}

\author[Datta,  Majumdar, Bharadwaj \& Choudhury  ]
{Kanan K. Datta$^1$\thanks{E-mail: kanan@cts.iitkgp.ernet.in}, 
Suman Majumdar$^1$\thanks{E-mail: suman1@cts.iitkgp.ernet.in},
Somnath Bharadwaj$^1$\thanks{E-mail: somnathb@iitkgp.ac.in}
and T. Roy Choudhury$^2$\thanks{E-mail: chou@ast.cam.ac.uk}\\
$^1$Department of Physics and Meteorology \&  
Centre for Theoretical Studies, IIT, Kharagpur 721302, India\\
$^2$Institute of Astronomy, Madingley Road, Cambridge CB3 0HA, UK}

\maketitle
\date{\today}

\begin{abstract}
Extending the  formalism of Datta, Bharadwaj \& Choudhury (2007) for
detecting  ionized bubbles in redshifted 21 cm maps using a 
matched-filtering technique, we use different simulations to 
analyze the impact of HI fluctuations outside the bubble on the
detectability of the bubble. In the first  three kinds of simulations
there is a 
spherical  bubble of comoving radius $R_b$, the one  that we are
trying to detect,   located at 
the center, and the  neutral hydrogen (HI) outside the bubble 
traces the underlying dark matter distribution. We consider 
 three different possible scenarios of reionization, i.e., (i) there
 is a single  bubble  (SB) in the field of view (FoV) and the
 hydrogen neutral fraction is constant outside this bubble 
 (ii) patchy reionization with many small ionized bubbles in the FoV
 (PR1)  and (iii) many spherical ionized bubbles of the same
radius $R_b$ (PR2).  The centers of the extra 
bubbles  trace the dark matter distribution. The fourth kind of
simulation uses  more realistic maps based on semi-numeric
modelling (SM) of ionized regions.
We make predictions
for the currently functioning GMRT and a forthcoming instrument, the
MWA at a redshift of 6 (corresponding to a observed frequency $203\,
{\rm MHz}$) for $1000 \,{\rm hrs}$ observations. We find that for both
the SB and PR1 scenarios the fluctuating IGM
restricts  bubble detection to size $R_b \le 6 \,
{\rm Mpc}$ and $R_b \le 12 \,  {\rm Mpc}$ for the GMRT and  the MWA
respectively, 
however large  be the integration time. These results are well
explained by analytical predictions. In the PR2 scenario, we find that
bubble detection is almost impossible for neutral fraction $x_{\rm
HI}<0.6$  because of large uncertainty due
to the HI fluctuations.  Applying the matched-filter  technique 
to the SM scenario, we find that it
works well even when the targeted ionized bubble is non-spherical due
to  surrounding bubbles and inhomogeneous recombination.
We find that determining the size and
positions of the bubbles is not limited  by the HI fluctuations in
the SB and PR1 scenario but limited  by the instrument's 
 angular resolution   instead, and this can be done more 
precisely for larger bubble. We also find that 
for  bubble detection the GMRT  configuration is somewhat superior to
the proposed MWA. 
\end{abstract}

\begin{keywords}
cosmology: theory, cosmology: diffuse radiation, Methods: data analysis
\end{keywords}

\section{Introduction}
The epoch of reionization is one of the least known chapters 
in the evolutionary history of the
Universe. Quasar absorption spectra \citep{becker,fan02,fan06a} and CMBR
observations \citep{spergel,page,dunkley08} together imply that
reionization 
occurred over an extended period spanning the redshift range $6 \le z
\le 15$ (for reviews see \citealt{fan06b,choudhury06}). It is believed that
the ionizing radiation from quasars and the stars within  galaxies 
reionize the surrounding neutral  intergalactic medium (IGM). Ionized bubbles
form around these  luminous objects,   grow and finally overlap
to completely reionize the Universe. 
The issue of  detecting  these  bubbles in 
radio-interferometric observations of redshifted 
HI $21$ cm radiation has been drawing considerable  attention. This is
motivated by the   Giant   Metre-Wave Radio Telescope 
 (GMRT\footnote{http://www.gmrt.ncra.tifr.res.in};  
\citealt{swarup}) which is currently functional 
and  several low frequency radio telescopes which
are expected to become functional in the future  
(eg. MWA\footnote{http://www.haystack.mit.edu/ast/arrays/mwa/},
LOFAR\footnote{http://www.lofar.org/}, 21 CMA
\footnote{http://web.phys.cmu.edu/$\sim$past/},
PAPER\footnote{http://astro.berkeley.edu/$\sim$dbacker/eor/}, VLA
extension program\footnote{http://www.cfa.harvard.edu/dawn/} and  
SKA\footnote{http://www.skatelescope.org/}). These  are all being
designed to  be sensitive  to the HI signal from 
the epoch of reionization. 

The  detection of individual ionized bubbles would be a direct probe
of the reionization process. It has been proposed  that 
such observations  will probe the   properties of the ionizing
sources and the evolution of the surrounding  IGM
\citep{wyithe04a,wyithe05,kohler05,maselli07,alvarez07,geil07,wyithe08,geil08}.
Observations of 
individual ionized bubbles would  complement the  study of
reionization through the power spectrum  of HI  brightness temperature
fluctuations \citep{zal04,morales04,bharad05,ali05,sethi05,datta1}.

Nearly  all the previous  work on detecting ionized regions consider 
the contrast between the ionized regions and the neutral IGM in images
of redshifted HI 21 cm radiation. The HI signal is expected to be
only a small contribution buried deep in the emission  from other
astrophysical sources (foregrounds) and in the system noise.  It is a
big challenge to detect the signal of an ionized bubble from the other
contributions that are orders of magnitude larger
\citep{shaver99,dimat1,oh03,cooray3,santos05,gleser07,ali08}. In an 
earlier work (\citealt{datta2}, hereafter referred to as Paper I ) we have
introduced a matched filter  to   optimally combine the
entire signal of an ionized  bubble while  minimizing the noise and
foreground  contributions. This technique uses the visibilities which
are the fundamental quantity measured in radio-interferometric
observations. Using visibilities  has an advantage  over the image
based techniques because the system noise contribution in different
visibilities is independent whereas   the noise in different pixels of
a radio-interferometric images is not.  

Paper I presents an analytic framework for predicting the expected 
value and the standard deviation $\sigma$ 
of the matched filter estimator for the detection of a 
 spherical ionized  bubble of comoving radius $R_b$. We identify three
 different contributions  to $\sigma$, namely
 foregrounds, system  noise and the fluctuations in the HI outside  
the bubble that we are trying to detect.  Our analysis shows 
that the matched filter  effectively removes the foreground  
contribution so  that it falls  below the signal. Considering the 
system noise for the GMRT and  the MWA  we find that a $3 \, \sigma$ detection 
will be possible  for a bubble  of comoving
radius $R_b \ge 40 \, {\rm Mpc}$  in $100 \, {\rm hrs}$ of observation
and $R_b \ge 22 \, {\rm Mpc}$ in $1000 \, {\rm hrs}$ of observation
for both the instruments. The HI fluctuations, we find, impose a
fundamental restriction on bubble detection. Under the assumption that
the HI outside the ionized bubble traces the dark matter we find that 
 it is not possible to detect bubble of
size $R_b \le 8 \, {\rm Mpc}$  and $R_b \le 16 \, {\rm Mpc}$ at the
GMRT and MWA respectively. Note that the matched filter technique 
is valid for both,  a targeted search around  QSOs as well as  for a blind
search in a random direction.

In this paper we validate the visibility based matched filter
technique introduced in Paper I through  simulations of bubble
detection. Our simulations are   capable of 
handling interferometric arrays with widely different 
configurations like the GMRT and the MWA , the two instruments that we
consider here.  As mentioned earlier, the  fluctuations in the HI
outside the target bubble impose a fundamental  restriction for bubble
detection. The analytic approach of Paper I assumes that the HI
outside the bubble traces the dark matter.  In this paper we
carry out simulations that incorporate this assumption and use these
to assess the impact of HI fluctuations for bubble detection. We also
use the simulations to determine the accuracy to which the GMRT and
the MWA will be able to determine the size  and the position  of an 
ionized bubble, and test if this is limited due to the presence of HI
fluctuations. 
In a real situation  a typical FoV is expected to  
contain  several ionized patches besides the one that we are trying to
detect. We  use simulations to assess the impact of HI fluctuations for
bubble detection  in patchy reionization scenarios.

The outline of the paper is as follows. Section  2 presents a 
brief description of how we simulate  21-cm maps for three different
scenarios of the HI distribution, one where the HI traces the dark
matter and two with patchy reionization.  Subsections 2.1 and 2.2
respectively discuss how the simulated maps are converted into
visibilities and  how the matched filter analysis is simulated.
We present our results in Section 3. Subsections 3.1, 3.2 and 3.3
present results for bubble detectability, size determination and
position determination under the assumption that the HI outside the
bubble traces the dark matter. Section 3.4 presents results for bubble
detectability in patchy reionization scenarios. We discuss redshift dependence 
of bubble detection in section 4 and present our summary
in section 5.

For the GMRT we have used the telescope parameters from their 
website, while for the MWA  we use the telescope parameters from
\citet{bowman07}. The cosmological parameters used
throughout this paper  are   
$\Omega_m=0.3, \Omega_b h^2 = 0.022, n_s = 1., h = 0.74, \sigma_8
= 1$.

\section{Method of Simulation}
We have simulated the detection of the HI signal of an ionized bubble
whose center is at
redshift $z_c=6$ which corresponds to $\nu_c=203 \, {\rm MHz}$. 
 The choice of $z$ value  is guided by the fact that we expect large 
ionized regions towards the end of reionization $z \gtrsim 6$
\citep{wyithe04b,furlanetto05}.  
Our aim here is to validate the analytic calculations of
Paper I and hence the exact value of $z$ is not very important.

We consider  four  scenarios of reionization for bubble detection. 
In the first three scenarios  there is  a spherical ionized bubble,
the one that we are trying to detect, at the center of the FoV. This 
bubble has comoving radius
$R_b$ and  is embedded in HI that traces the dark matter. 
In the first scenario  there is  a single bubble  in the field 
of view. We refer to this  as the SB scenario. In this scenario the
HI fraction $x_{\rm HI}$ is assumed to be 
uniform outside the bubble. The uncertainty due to the HI
fluctuations is expected to be lowest  
in this scenario because of the absence of patchiness. This is the
most optimistic scenario  for bubble detection. 
\begin{figure*}
\includegraphics[width=0.25\textwidth, angle=270]{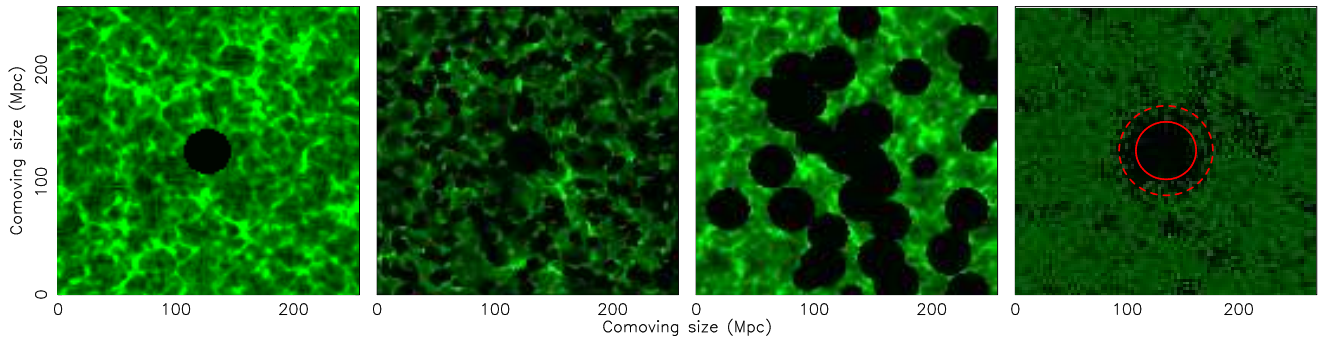}
\caption{This shows HI images on slices through the center of the
  bubble for the  four different
  scenarios SB, PR1, PR2 and SM (from left to right). In first three
  panels  the 
  central,  circular dark region of radius $R_b=20 \, {\rm Mpc}$ shows
  the 
  HII bubble  that we are trying to detect. The  HI outside this bubble 
  traces the dark  matter distribution. In the SB scenario(left)
  the hydrogen neutral 
  fraction is $x_{\rm HI}=1$  outside the bubble. In the 
  PR1  scenario (2nd from left) the extra    bubbles are all of a fixed
  comoving  radius $6 \, {\rm Mpc}$. In the  PR2  scenario (3rd from
  left) the   extra   bubbles have the same   comoving radius as the
  bubble that   we are trying to detect.   In both the   PR1 and
  PR2 reionization scenarios 
  the centers of the extra  bubbles trace the dark matter distribution
  and  $x_{\rm    HI}=0.62$. In the SM scenario (right) the central
  region up to radius $27\, {\rm Mpc} $ is fully ionized (marked with
  solid circle) and beyond that region up to radius $42\, {\rm Mpc} $
  the region is partially filled with HI patches (dashed circle). The
  mean neutral fraction is $x_{\rm HI}=0.5$. These simulations are
  all   for the   GMRT.} 
\label{fig:image}
\end{figure*}

In the next  two scenarios, we  attempt to quantify the effect of
patchy reionization (PR)  outside the bubble that we are trying to detect 
by introducing many other, possibly overlapping, bubbles in the FoV.  
Unfortunately, there is no obvious way to fix the sizes of these bubbles
from any theoretical models as they depend crucially on the nature of
reionization sources and other physical factors. 
In scenario PR1, we assume that the large HII
regions which we are trying to detect are surrounded by many
small ionized regions whose sizes are fixed by
the following procedure: we assume the globally averaged
neutral fraction $x_{\rm HI}$ to be $\sim 0.5$; the reason for this
choice is that the effects of patchiness
would be most prominent when typically half of the IGM is ionized. 
Given the value of $x_{\rm HI}$, we try to obtain a reasonable 
estimate of the size of the background bubbles from available
models. For example, semi-numeric simulations of patchy reionization  
\citep{mesinger07} predict that  the bubble size distribution  peaks around
$5 \,{\rm Mpc}$ when $x_{\rm HI} = 0.61$ (see their Fig 6).
We thus choose the spherical background bubbles to have radii $6
\,{\rm Mpc}$ and compute 
the number of 
background bubbles by demanding that the resulting neutral fraction
is 0.5. The bubble centres are chosen such that they trace the 
underlying dark matter distribution. At the end, the value of 
$x_{\rm HI}$ turns out to be slightly higher 0.62 because of overlap of the 
bubbles. Note that because of these overlaps, 
the shapes of the resulting ionized regions would not always be 
perfectly spherical.
In this scenario, we have essentially attempted
to capture a situation where there are many small, possibly
overlapping  ionized regions produced by galaxies and a few large
ionized regions (like the one that we are trying to detect) produced
by QSOs. 

Since the choice of the background bubble size is not robust 
by any means,
we consider a different scenario PR2  where these bubbles have the same 
comoving radius $R_b$ as the bubble that we are trying to detect. 
The centers of these extra
bubbles trace the dark matter distribution as in PR1.
The number of bubbles is fixed by the globally averaged 
$x_{\rm HI}$ which we take to be 0.62 same as in PR1.
The PR2 scenario represents a situation where we
predominantly  have  large  ionized regions produced 
either by rare luminous sources or
through the overlap  of several small ionized regions in the later
stages of reionization.

A  particle-mesh (PM) N-body code   was used to simulate the dark
matter distribution.  Our earlier work (Paper I) shows that the
HI signal of the ionized bubble is largely concentrated at small
baselines or large angular scales,  thus a very high
spatial resolution is not required. 
We have used a  grid spacing of $2 \,{\rm Mpc}$  for the simulations. 
This is adequate for bubbles in the range 
 $4 \le R_b \le 50 \, {\rm Mpc}$
that we consider. The simulations use 
 $256^3$ particles on a $256^3$ mesh. For the 
GMRT  a single N-body simulation was cut into $8$ equal cubes  of size
$256 \, {\rm 
  Mpc}$ on each side. Considering that each cube may be viewed 
along three different directions, we have a total of $24$ different
realization of the dark matter distribution. Each cube  corresponds to
$18 \, {\rm MHz}$ in frequency and 
$\sim 2\degr$ in angle which is comparable to the GMRT FoV
 which has FWHM=$1.7\degr$ at $203 {\rm MHz}$.  The MWA FoV is
much larger (FWHM=$13\degr$). Here eight independent N-body 
simulations were used.  Viewing these along three different directions
gives twenty four different realizations of the dark matter
distribution. Limited computer memory restricts the simulation size 
and   the angular extent  $(\sim 4^{\degr})$ is considerably smaller
than the MWA FoV. We do not expect this to affect the signal but the
contribution from the HI fluctuations  outside the bubble is
possibly underestimated for the MWA.

The dark matter density contrast $\delta$ was used to calculate the 
redshifted $21-{\rm cm}$ specific intensity $I_{\nu}=\bar{I}_{\nu}
x_{\rm HI} (1+\delta)$   for each grid point of our simulation.
Here $\bar{I}_{\nu}=2.5\times10^2\frac{Jy}{sr} \left (\frac{\Omega_b 
h^2}{0.02}\right )\left( \frac{0.7}{h} \right ) \left
(\frac{H_0}{H(z)} \right ) $ and $x_{\rm HI}$ the hydrogen neutral
fraction is $0$ inside the ionized bubbles and $1$ outside. 
The simulated boxes are transformed to frequency and sky coordinate. 
Figure \ref{fig:image} shows  the HI image on a slice through the
center of the bubble of radius $R_b=20 \, {\rm Mpc}$.  The mean 
neutral fraction $\bar{x}_{\rm HI}$
is $\sim 1$ in the SB scenario, while it is $\sim
0.62$ for the two PR simulations shown here.

The three scenarios discussed above consider only spherical bubbles,
and the  only departures from sphericity arise from bubble
overlap. It is important to  assess how well  
 our bubble detection technique works   for non-spherical bubbles,   
which we do using  ionization maps produced by the semi-numeric (SM) 
approach.  In particular, we use maps obtained by the method of
\cite{choudhury08}. 
Essentially, these maps are produced by incorporating an excursion-set
based technique for identifying ionized regions given the density
distribution and the ionizing sources \citep{zahn07,mesinger07,geil07}.
In addition, the method of \cite{choudhury08} incorporate
inhomogeneous recombination and self-shielding of high-density regions
so that it is consistent 
with the ``photons-starved'' reionization scenario implied
by the Ly$\alpha$ forest data \citep{bolton07,choudhury08a}.
We use a simulation box of size 270 Mpc with $2000^3$ particles
which can resolve collapsed halos as small as $\approx 10^9 M_{\odot}$.
The ionization maps are generated at a much lower resolution
with a
grid size of $2.7$ Mpc. The box 
corresponds to $19\, {\bf MHz}$ in frequency and $\sim2\degr$ in angle 
comparable to the GMRT FoV. We have assigned luminosities to the collapsed
halos such that the mean neutral fraction $x_{\bf HI}=0.5$. The most massive
halo (mass $\sim 10^{13} M_{\odot}$) identified  in the box 
is made to coincide with the 
box centre and we assume that
it hosts a luminous QSO; its luminosity and age are chosen 
such that it would produce a 
spherical HII region of comoving size $\approx$ 27 Mpc in a completely homogeneous neutral medium [see, e.g., equation (8) of \cite{geil07}]. However, the actual ionized region is far from spherical both
because of the surrounding bubbles from other halos and also
because of inhomogeneous recombination.
We find visually from the maps (see the rightmost panel of Figure
\ref{fig:image}) that the HII region is fully ionized up to radius
$\approx 27$ Mpc. Beyond that the region is partially filled with
neutral patches. This patchy ionized region extends up to radius $\sim
42$ Mpc and then merges with the average IGM.  The fully ionized
region and the region with HI patches are marked with two circles. We
use this box for GMRT as three independent realizations viewing the
box along three different directions.  For the MWA we need a much
larger simulation box which requires substantially more 
computing power, beyond the resources available to us at present. 
Hence we do not consider the MWA for this scenario. 

\subsection{Simulating Visibilities}
The quantity measured in radio-interferometric 
observations is the visibility $V(\u,\nu)$ which is related to the specific
intensity pattern on the sky $I_{\nu}(\th)$ as 
\be
V(\u,\nu)=\int d^2 \theta A(\th) I_{\nu}(\th)
e^{ 2\pi \imath \th \cdot \u}
\label{eq:vis}
\e
Here the baseline $\u={\vec d}/\lambda$ denotes 
the antenna separation ${\vec d}$  projected in the plane 
perpendicular to the line of sight  in units of the observing
wavelength $\lambda$, $\th$ is a two dimensional vector in the plane
of the sky with
origin at the center of the FoV, and $A(\th)$ is the 
beam  pattern of the individual antenna. For the GMRT this can be well 
approximated by Gaussian $A(\th)=e^{-{\theta}^2/{\theta_0}^2}$ where 
$\theta_0 \approx 0.6 ~\theta_{\rm FWHM}$ and we use the values
$1.7\degr$  for $\theta_0$  at $203\, {\rm MHz}$ corresponding to
the redshift $z=6$ for the GMRT. The MWA beam pattern is expected
to be quite complicated, and depends on 
the pointing angle relative to the zenith \citep{bowman07}. Our analysis
largely deals with the beam pattern within $2^{\circ}$ of the pointing
angle where it is reasonable to approximate the beam as being
circularly symmetric (Figures  3 and 5 of \citealt{bowman07} ). We  
approximate the MWA antenna beam pattern as a Gaussian. 

We consider  $128$ frequency channels across  $18 {\rm MHz}$
bandwidth. The image $I_{\nu}(\theta)$ at each channel is multiplied
with  the   telescope beam pattern $A(\vec {\theta},\nu)$.  The
discrete Fourier transform (DFT) of the product $I_{\nu}(\theta) \,
A(\vec {\theta},\nu)$  gives the 
complex visibilities $\hat V(\vec{U},\nu)$. The GMRT simulations 
have baselines in the range $30.5 \le U \le 3900$
which is adequate to capture the HI  signal from  ionized bubbles
which  is expected to be confined to  small baselines $U<1000$. 

\begin{figure*}
\includegraphics[width=0.45\textwidth, angle=270]{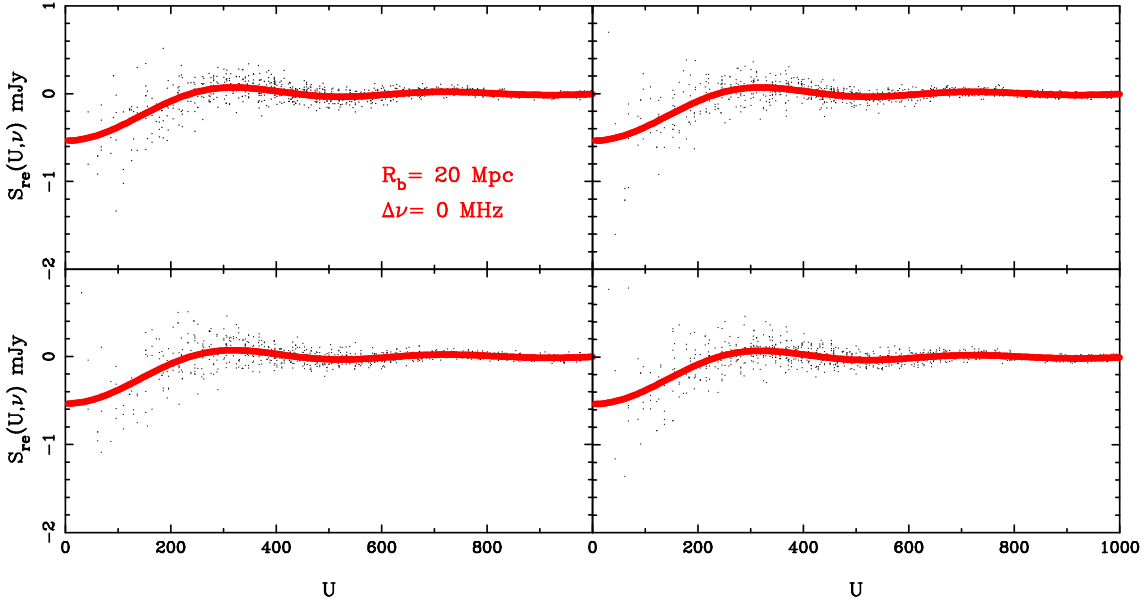}
\caption{This shows the visibility signal (real part) from a frequency
  slice   through the   center of a spherical ionized bubble 
  of comoving radius $20 \, {\rm Mpc}$ embedded in HI. 
  The solid  curves show the expected signal assuming that the bubble is
  embedded in uniformly   distributed HI. The data points show
  the visibilities for a few randomly chosen baselines from our
  simulation of the SB scenario.  The difference between the data
  points and  the solid curve
  is due to the fluctuations in the HI outside the bubble.  Each panel
  corresponds to a different   realization of the simulation. }  
\label{fig:vis_U}
\end{figure*}
\begin{figure*}
\includegraphics[width=0.53\textwidth, angle=270]{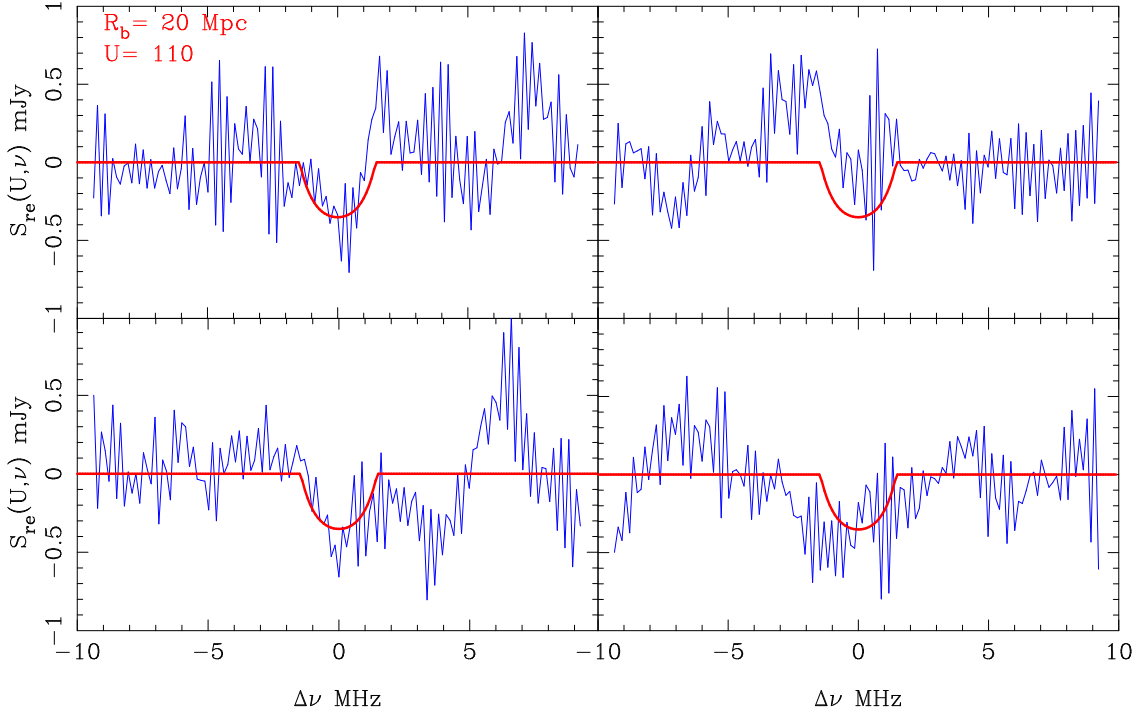}
\caption{Same as the previous figure except that $U$ is fixed at $110$
  while  the frequency varies , and  $\Delta \nu=\nu-\nu_c$.}
\label{fig:vis_nu}
\end{figure*}

The visibility  recorded in radio-interferometric
observations is actually a combination of  several  contributions 
\be
V(\vec{U},\nu)=S(\vec{U},\nu)+HF(\vec{U},\nu)+N(\vec{U},\nu)+F(\vec{U},\nu)
\,. 
\label{eq:2}
\e
The signal  $S(\u,\nu)$ from an ionized bubble of comoving radius
$R_b$ embedded in an uniform HI distribution  can be analytically
calculated (Paper I). The solid curve in Figures \ref{fig:vis_U} and
\ref{fig:vis_nu}  show the expected signal for $R_b=20 \, {\rm
  Mpc}$.  
The $U$ extent,  frequency extent and peak value of the signal scale
as  $ R_b^{-1}$, $R_b$ and $R_b^2$ respectively for other values of 
$R_b$.    Note that   $S(\u,\nu)$ is real when the bubble is at the 
center of the FoV.

The data points shown  in Figures \ref{fig:vis_U} and \ref{fig:vis_nu}  
are the real part of a few randomly chosen visibilities determined
from the simulation of a  $R_b=20 \, {\rm Mpc}$ bubble in the
SB scenario. The deviations from the analytic predictions are due to
the HI fluctuations  $HF(\u,\nu)$  {\it ie.} in the SB scenario the HI
outside the bubble traces the dark matter fluctuations. Notice that
these fluctuations  are often so prominent that the signal cannot be
made out.   We expect even larger fluctuations  in the other three scenarios which incorporate patchiness of reionization.

The system noise contribution $N(\u,\nu)$ in each baseline and
frequency channel is expected to be an independent Gaussian random
variable with zero mean ($\langle \hat{N} \rangle =0$) and  
variance $\sqrt{\langle \hat{N}^2  \rangle}$is independent of   $\u$
and $\nu_c$. We use (Paper I)  
\be
\sqrt{\langle \hat{N}^2  \rangle}
=C^x \left (\frac{\Delta \nu_c}{1
   {\rm MHz}} \right )^{-1/2}\left ( \frac{\Delta
    t}{1 \rm{sec}}\right )^{-1/2}
\label{eq:noise}
\e
where $C^x$ has values $0.53{\rm Jy}$
and $54.21{\rm Jy}$ for the GMRT and the MWA respectively (Paper I).

The contribution from astrophysical foregrounds $F(\vec{U},\nu)$ is
expected  to be several order of magnitude stronger than the HI
signal.   The foregrounds are predicted to have a featureless,
continuum spectra 
whereas the signal is expected to have a dip 
 at $\nu_c$ (Figure \ref{fig:vis_nu}). This difference holds the
 promise of allowing us to separate the signal from the foregrounds. 
\begin{figure*}
\includegraphics[width=0.32\textwidth, angle=270]{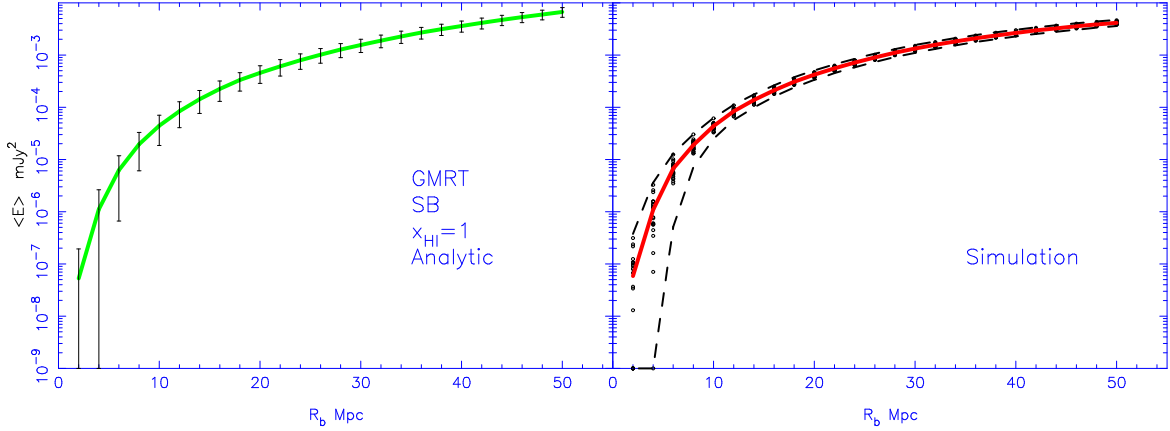}
\caption{The estimator $\hat{E}$ (defined in equation
  \ref{eq:estim0}) for bubble size
  $R_b$ ranging from $4\,{\rm Mpc}$ to $50\,{\rm Mpc}$ for the GMRT in
  the SB  scenario.  It is assumed that the filter is exactly matched
  to the  bubble. The left panel shows the analytic predictions for
  the  mean  estimator 
  $\langle\hat{E}\rangle$  and the $3-\sigma$ error-bars
  due to the HI fluctuations. The solid and  the dashed lines 
in the right panel respectively  show the   $\langle\hat{E}\rangle$
  and the   $3-\sigma$  envelope determined from  the simulations.  
The data points in the right panel show  $\hat{E}$ in the individual
  realizations.  
}
\label{fig:R_Es}
\end{figure*}

\begin{figure*}
\includegraphics[width=0.32\textwidth, angle=270]{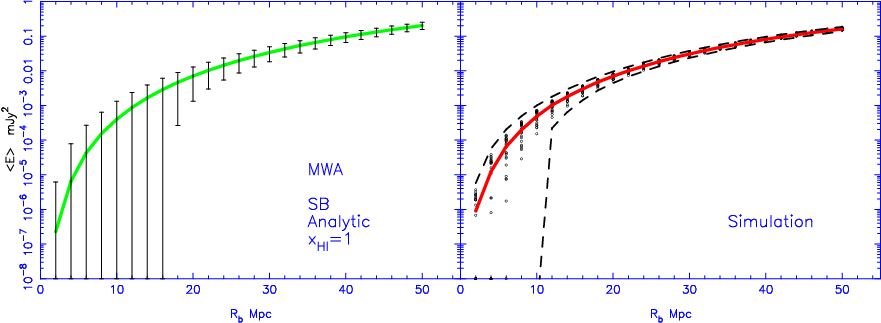}
\caption{Same as the Figure \ref{fig:R_Es} for the MWA.}
\label{fig:R_EsM}
\end{figure*}

\subsection{Simulating Signal Detection}

The signal component $S(\u,\nu)$  in the observed visibilities 
$V(\u,\nu)$ is expected to be  buried deep in other 
contributions many of which are orders of magnitude larger. Detecting
this is a big challenge. For optimal signal detection 
we consider the  estimator (Paper I) 
\be
\hat{E}= \sum_{a,b} S_{f}^{\ast}(\u_a,\nu_b)
\hat{V}(\u_a,\nu_b) 
\label{eq:estim0}
\e
where 
$S_f(\u,\nu)$ is a filter which has been constructed to detect
a particular ionized bubble, $\hat{V}(\u_a,\nu_b)$ refer to the
observed visibilities and  $\u_a$ and $\nu_b$ refer to the
different baselines and frequency channels in the observation. 
The filter $S_f(\u,\nu)$  depends on $[R_f,z_c,\th_c]$
the  comoving  radius, redshift and angular position of the bubble
that we are trying to detect.  We do not show this  
explicitly,  the values of these parameters will be clear
from the context.

The  baselines  obtained using DFT in our simulations are uniformly
distributed on a plane. In real observations, the baselines will have
a complicated distribution depending on the antenna layout and
direction of observation. We incorporate this through the normalized
baseline distribution function $\rho_N(U,\nu)$  which  is defined such that 
$d^2 Ud\nu \, \rho_N(\u,\nu)$ is the fraction  of data points {\it
  ie.} baselines and  frequency channels in the
interval  $d^2 U \, d \nu$ and $\int d^2 U \,\int d\nu \,
\rho_N(\u,\nu)=1$.  
We use the functional forms of   $\rho_N$ determined in Paper I  for
the GMRT and the MWA. 

Using  the  simulated visibilities, we evaluate the estimator 
as
\be
\hat{E}= \, (\Delta U)^2 \, \Delta \nu \, 
 \sum_{a,b} S_{f}^{\ast}(\u_a,\nu_b)
\hat{V}(\u_a,\nu_b) \rho_N(\u_a,\nu_b) 
\label{eq:estim1}
\e
where the sum is now over the baselines and frequency channels in the
simulation.

The filter $S_f(\u, \,\nu)$ (Filter I of  Paper I)  is defined as 
\bear
S_f(\u,
\,\nu)\!\!\!\!\!&=&\!\!\!\!\!\left(\frac{\nu}{\nu_c}\right)^2\left[
  S(\u, \,\nu)\right. 
\nline
\!\!\!\!\!&-&\!\!\!\!\! \left. \frac{\Theta(1-2 \mid \nu -\nu_c
  \mid/B^{'}) }{ B^{'}} 
\int_{\nu_c-B^{'}/2}^{\nu_c + B^{'}/2}S(\u,\nu') \, \de \nu' \right
]. \nonumber\\ 
\label{eq:rms_fg}
\ear
where the first term $ S(\u, \,\nu)$ is the expected signal of the
bubble that we are trying to detect. We note that this term is the
matched filter that gives the maximum signal to noise ratio (SNR).  The
second term involving the Heaviside function $\Theta(x)$ 
subtracts out any frequency independent  component from  the frequency
range $\nu_c-B^{'}/2$ to $\nu_c+B^{'}/2$. The latter term is
introduced to subtract out the foreground contributions. The 
$(\nu/\nu_c)^2$ term accounts for the fact that
$\rho_N(U,\nu)$ changes with frequency (equivalently wavelength).

We have used the $24$  independent realizations
of the simulation   for the first three scenarios to  determine the
mean  $\langle\hat{E}\rangle$ and 
the  variance $\langle(\Delta\hat{E})^2\rangle$ of the 
estimator. The high computational requirement restricts us to use 
just  $3$ realizations  for the SM scenario.
 Only the
  signal is correlated with the filter, and only this  is 
expected to contribute to the  mean  $\langle\hat{E}\rangle$. All the
other components are uncorrelated with the filter and they are
expected to contribute only to the variance
$\langle(\Delta\hat{E})^2\rangle$. 
 The variance is a sum of three contributions (Paper I)
\begin{equation}
 \langle(\Delta\hat{E})^2\rangle= \langle(\Delta\hat{E})^2\rangle_{HF} +
 \langle(\Delta\hat{E})^2\rangle_{N} +
 \langle(\Delta\hat{E})^2\rangle_{FG} \,.
\end{equation}
 The simulations give an estimate of
 $\langle(\Delta\hat{E})^2\rangle_{HF}$  the contribution from HI
 fluctuations.  We do not include system noise explicitly 
in our  simulations. The noise contribution from a single visibility 
 (eq. \ref{eq:noise})   is used  to estimate 
 $ \langle(\Delta\hat{E})^2\rangle_{N}$ (eq. 19 of Paper I).  Under the assumed foreground
 model, the foreground contribution $
 \langle(\Delta\hat{E})^2\rangle_{FG} $ is predicted to be smaller
 than the signal and we do not consider it here.

\section{Results}

We first  consider the detection  of an ionized bubbles and the
estimation  of its parameters in the SB scenario where there is only 
a single bubble in the FoV. We consider the most optimistic situation
where the bubble is located in the center. 
 In reality this can  only be achieved in targeted observations of 
 ionized bubbles around luminous QSOs. In a blind search, the bubble
 in general will be located at some arbitrary position in the FoV, and
 not the center.  It has already been mentioned that the  
foregrounds can be removed by a suitable choice of the filter. 
Further, the system noise can, in principle,   be  reduced by
increasing the observation time. The HI fluctuations outside the
bubble impose a fundamental restriction on bubble detection. 

\subsection{Restriction  on bubble detection}
We have carried out simulations for different values of the  bubble
radius $R_b$ chosen uniformly at an interval of $2 \, {\rm Mpc}$
in the range $4$ to $50 \, {\rm Mpc}$.  In each case 
we consider only the most  optimistic situation where 
the bubble radius $R_f$ used in the filter is precisely
matched to   $R_b$. In reality  it is necessary to try filters of
different radius  $R_f$ to determine which gives the best match.  

Figures \ref{fig:R_Es} and  \ref{fig:R_EsM} 
shows the results  for the GMRT and the MWA
respectively.   We  compare the analytic predictions of Paper I (left 
panel) with  the prediction of our simulations (right
panel). The  analytical predictions for  the mean value 
$\langle\hat{E}\rangle$ arising from the signal 
 and 
$  \sqrt{\langle(\Delta\hat{E})^2\rangle}_{HF}$  due to the HI 
fluctuations  are respectively calculated using equations (15) and
(22) of Paper I. The  signal 
depends on the bubble radius $R_b$ and the mean neutral fraction
which is taken to be $x_{\rm HI}=1$.  The uncertainty due to the
HI fluctuations is calculated  using the  dark matter power spectrum
under the assumption that the HI traces the dark matter.

We find that $\langle\hat{E}\rangle$ and
$\sqrt{\langle(\Delta\hat{E})^2\rangle}_{HF}$ determined from the
simulations is in rough agreement with the analytic predictions.  
The mean $\langle\hat{E}\rangle$ is in very good agreement 
for $R_b>6 \, {\rm Mpc}$, there is a slight  discrepancy for smaller
bubbles arising from  the finite grid size $(2 \, {\rm
  Mpc}$ in  the simulation). 
The HI fluctuations $\sqrt{\langle(\Delta\hat{E})^2\rangle}_{HF}$ are  
somewhat  underestimated by the simulations. This is  more
pronounced for the MWA  where the limited box size of our  simulations
results in a   FoV  which is considerably smaller than the actual
antennas.  We note that the  $24$ different
values of $\hat E$ determined from the different realizations of the
simulation  all lie within $\langle\hat{E}\rangle \pm 
3 \sqrt{\langle(\Delta\hat{E})^2\rangle}_{HF}$  determined from the
analytic predictions. 

The good agreement between the simulation
results and the analytical predictions  is particularly important
because each  is based on several approximations,
many of which  differ between  the two methods.  Our results
show that the effect of these approximations, though present, 
are well under control. The  analytical method has the advantage 
that it is very   easy to  calculate and  can be evaluated 
very quickly at an extremely low computational cost. 
 Unfortunately, its utility is mainly  limited to the
SB scenario and it cannot be easily applied to an arbitrary PR scenario
with a complicated  HI  distribution. Simulations, though
computationally more cumbersome and expensive, are useful  in such a
situation. It is thus important to test that the two methods 
agree for the SB scenario where both of them  can be
applied. Note that the HI fluctuation predicted by the  
SB scenario sets the lower limit for the HI fluctuation in any of the
PR models. It is expected that patchiness will increase the HI
fluctuations above the SB predictions. 

\begin{figure*}
\includegraphics[width=0.4\textwidth, angle=270]{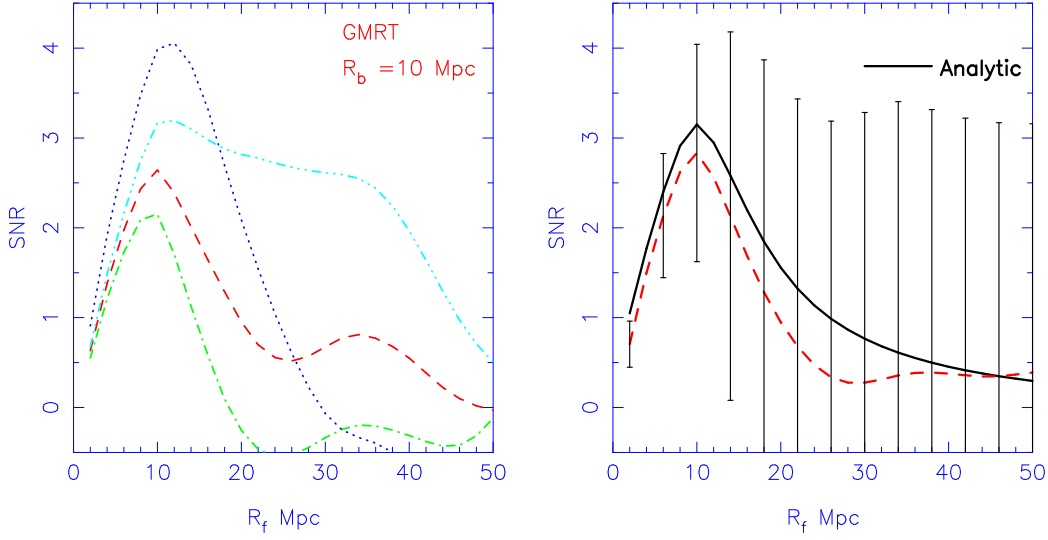}
\caption{The SNR $=\langle\hat{E}\rangle/\sqrt{\langle (\Delta \hat
    E)^2 \rangle_{\rm NS}}$ for  $1000\,{\rm
  hrs}$ observation with the GMRT 
as a function of the filter size $R_f$ for the
  case where the actual  bubble size is $R_b=10 \, {\rm Mpc}$ . The left panel shows $4$ different realizations 
    of the simulation.  The right panel shows the mean SNR and
    $3-\sigma$ error-bars
    calculated using 24 realizations. The solid line
    shows the analytical predictions. }
\label{fig:find_R10}
\end{figure*}
\begin{figure*}
\includegraphics[width=0.4\textwidth, angle=270]{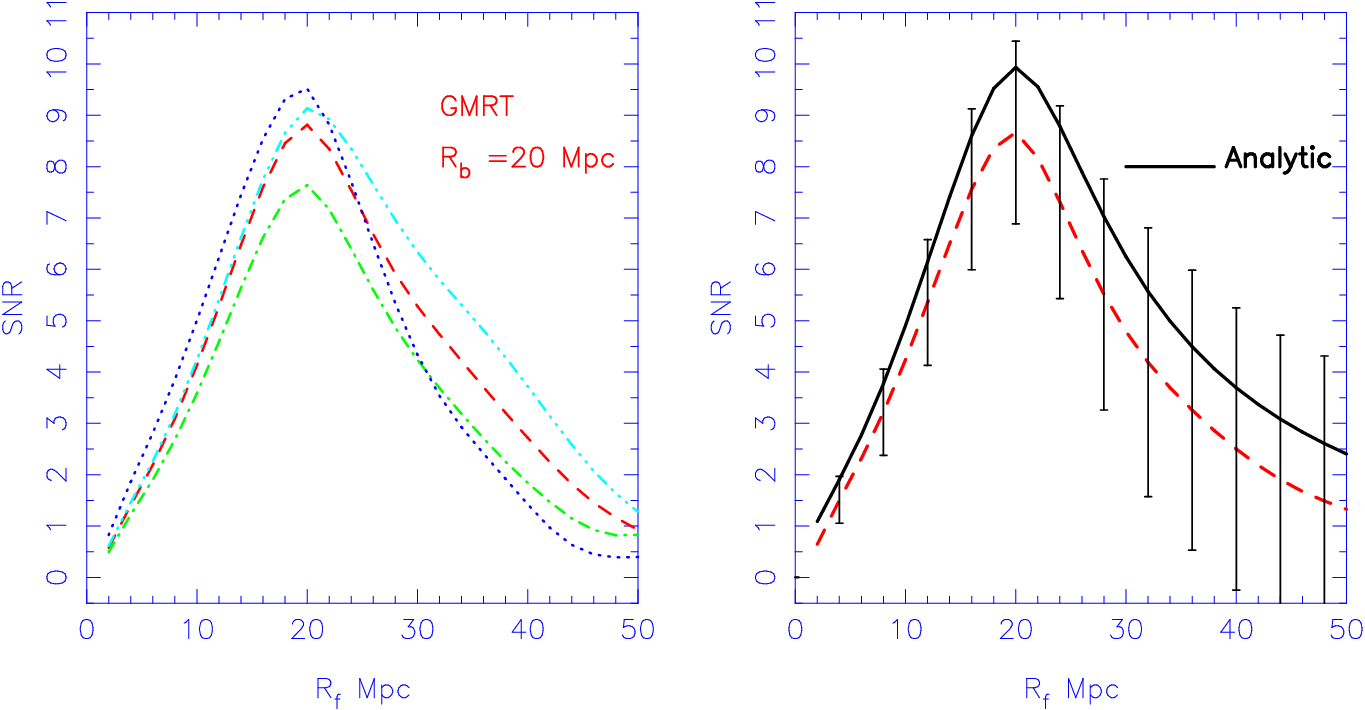}
\caption{Same as the Figure \ref{fig:find_R10} for
$R_b=20 \, {\rm Mpc}$  for the GMRT.}
\label{fig:find_R20}
\end{figure*}

\begin{figure*}
\includegraphics[width=0.4\textwidth, angle=270]{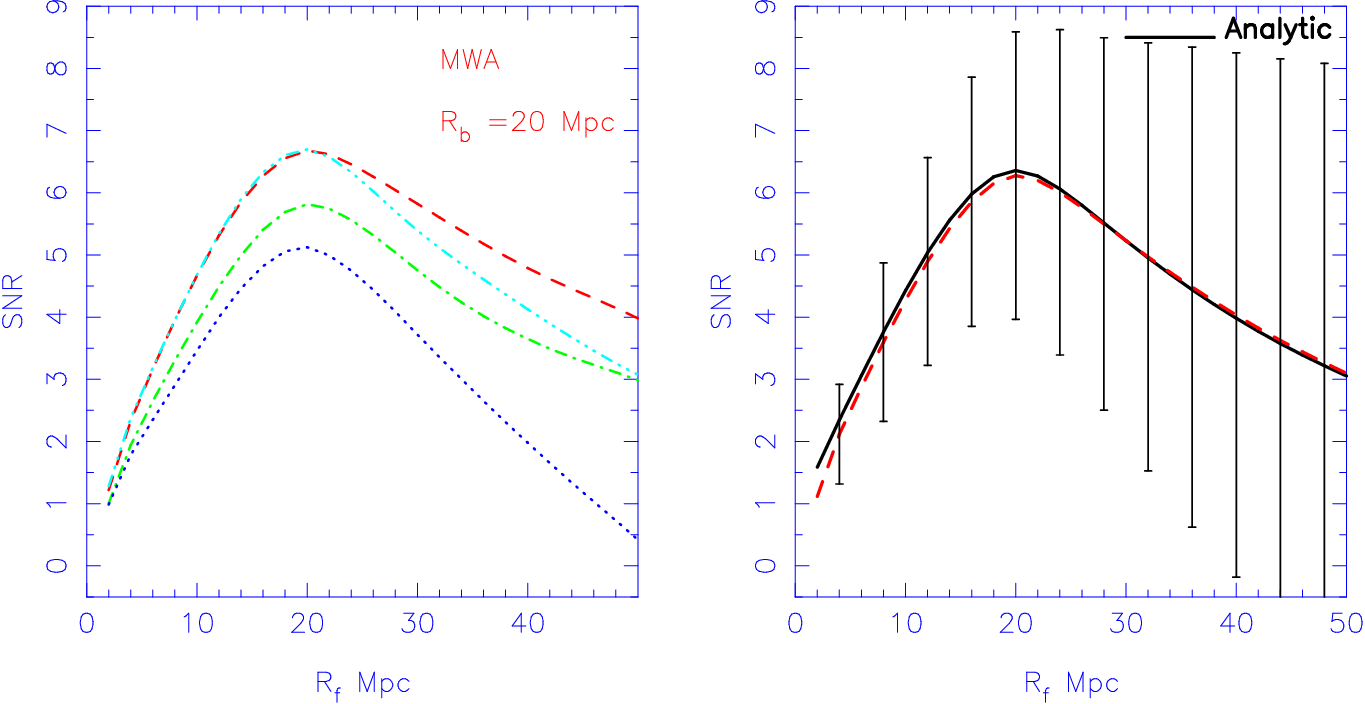}
\caption{Same as the Figure \ref{fig:find_R10}  for
$R_b=20 \, {\rm Mpc}$  for the MWA.}
\label{fig:find_R20M}
\end{figure*}

\begin{figure*}
\includegraphics[width=.65\textwidth, angle=270]{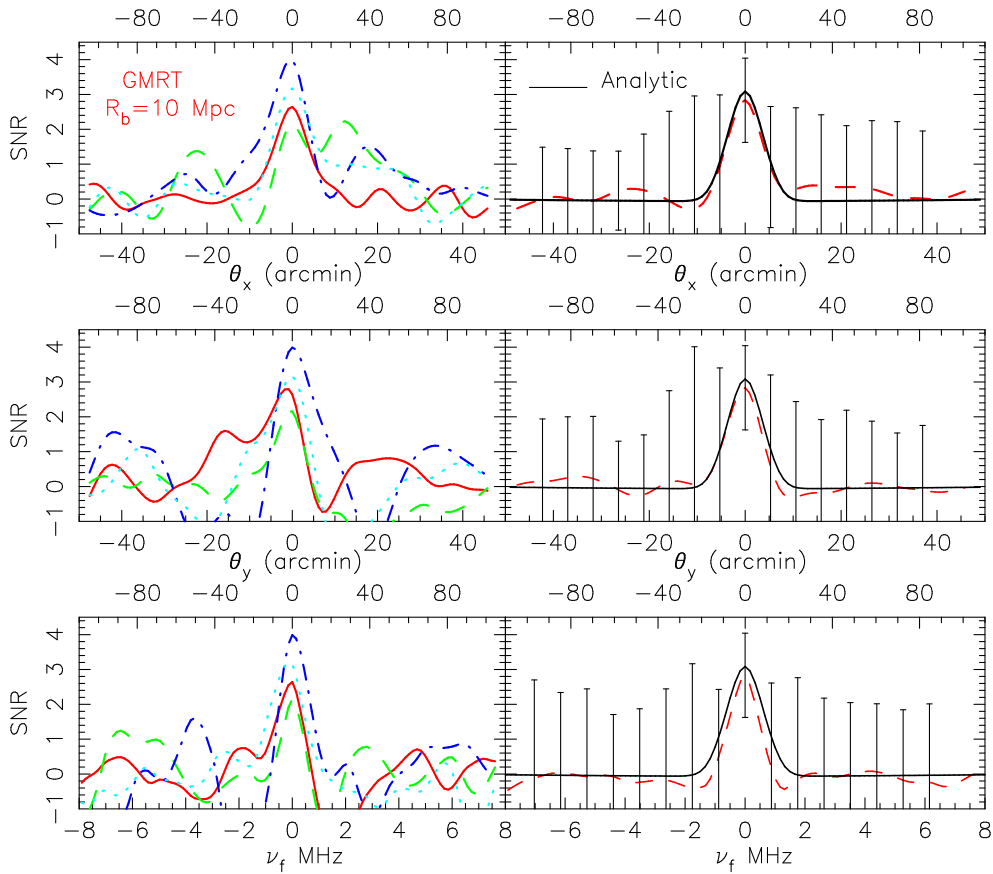}
\caption{The  SNR for $1000 \, {\rm hrs}$ GMRT observations 
for a   bubble of   size $R_b=10 \, {\rm Mpc}$
  located at the  center of the filed of view. The 
  filter  scans along  
  $\theta_x,\theta_Y,\nu_f$ (top, middle, bottom) to determine the
  bubble's position. 
 The left panel
  shows results for $4$  realizations of the SB simulation, the right
  panels show the mean (dashed curve) and $3-\sigma$   error-bars
  determined from $24$   realizations of the simulation and the
  analytic prediction  for the mean (solid curve).}
\label{fig:find_theta_nu10}
\end{figure*}

\begin{figure*}
\includegraphics[width=.65\textwidth, angle=270]{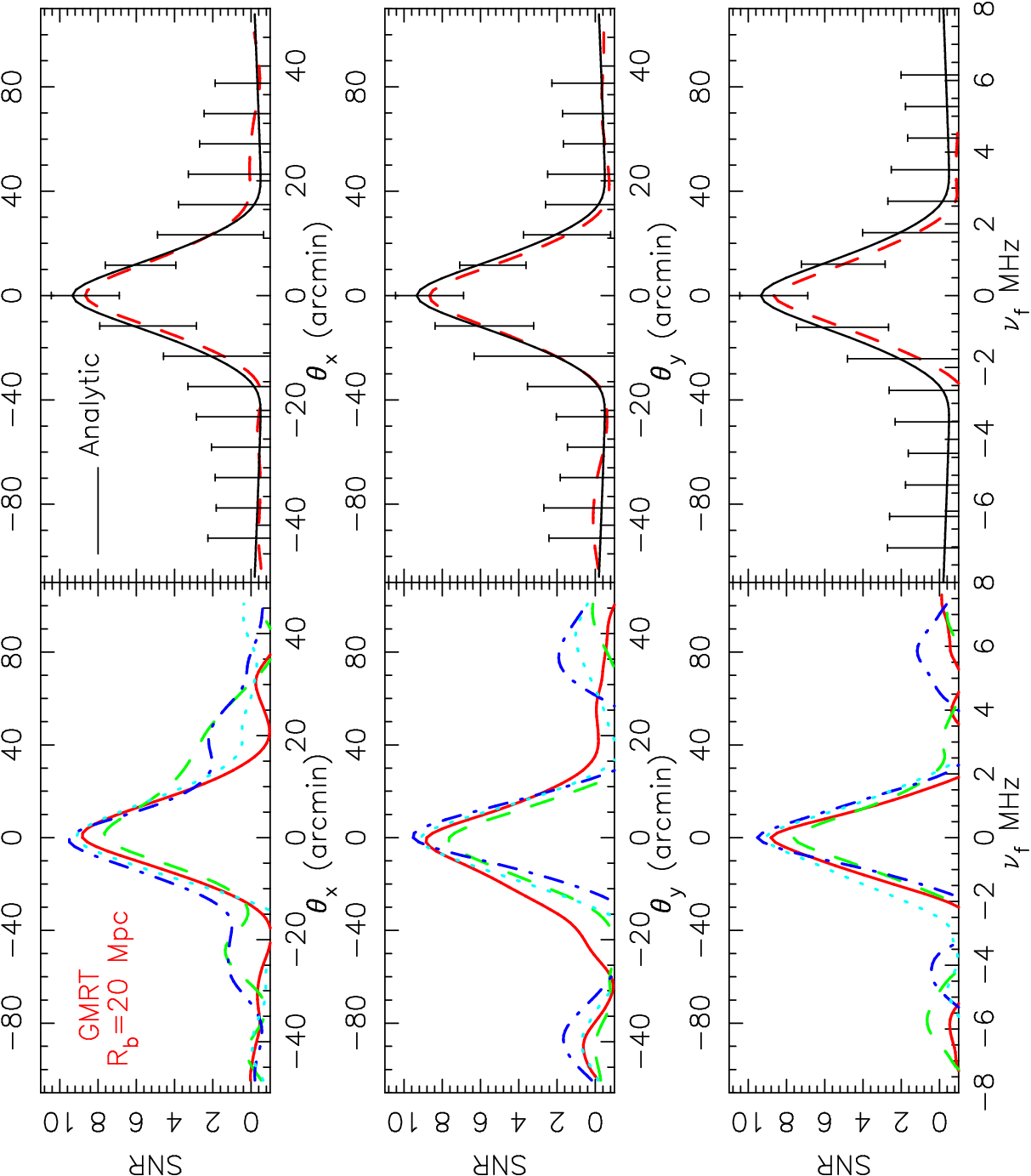}
\caption{Same as  the Figure \ref{fig:find_theta_nu10} for 
  $R_b=20 \, {\rm Mpc}$ for the GMRT.}
\label{fig:find_theta_nu20}
\end{figure*}

\begin{figure*}
\includegraphics[width=.65\textwidth, angle=270]{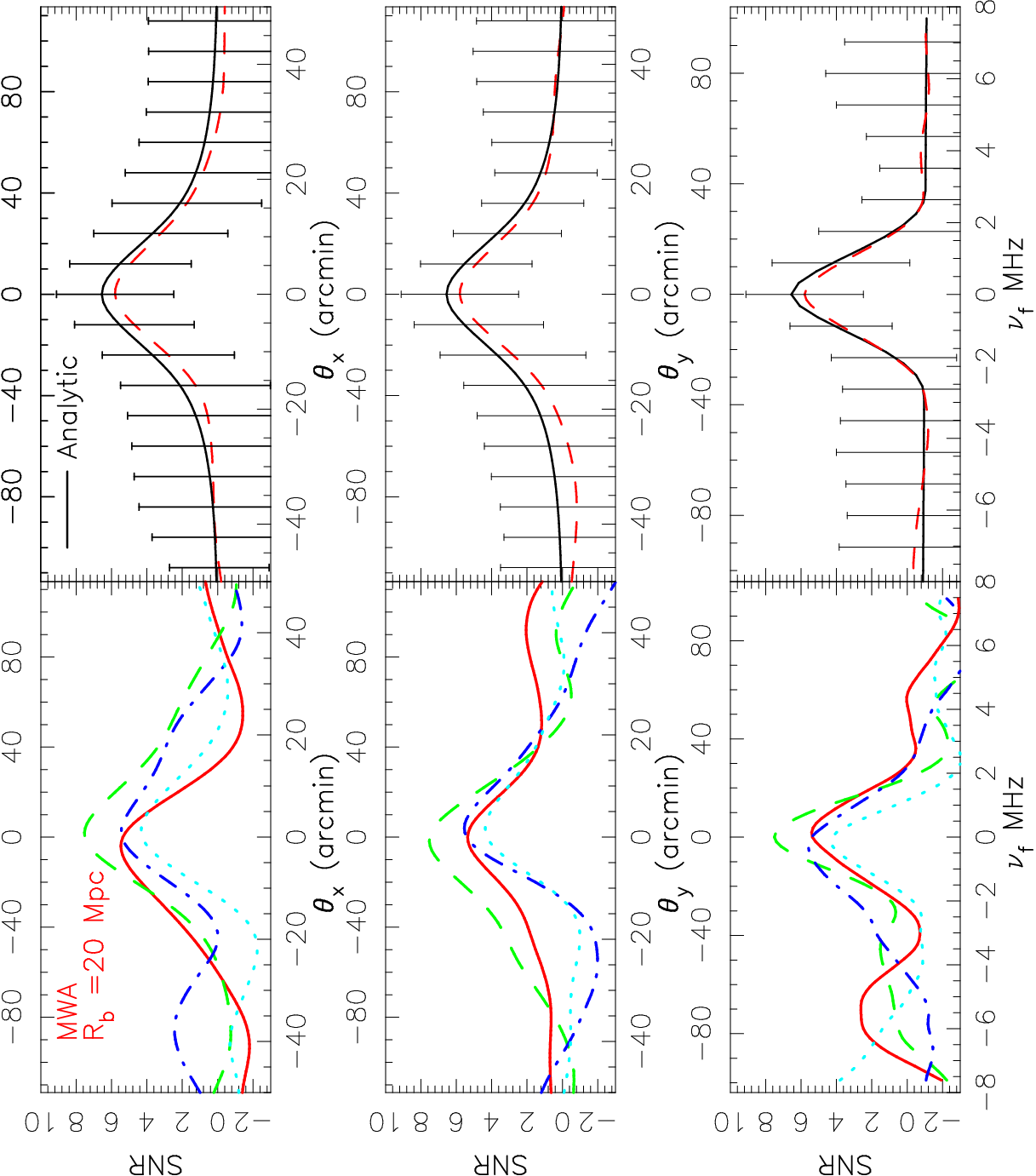}
\caption{Same as  the Figure \ref{fig:find_theta_nu10} for 
  $R_b=20 \, {\rm Mpc}$ for the MWA.}
\label{fig:find_theta_nu20M}
\end{figure*}

It is meaningful to attempt bubble detection at, say $3\sigma$
confidence level, only if  $\langle \hat{E} \rangle \ge 3
\sqrt{\langle (\Delta \hat{E})^2   \rangle_{HF}}$. The HI fluctuations
overwhelm the signal in a  situation where this condition is not 
satisfied, and bubble detection is not possible. 
In a  situation where this condition is satisfied, an
observed value $E_o$ of the estimator can be interpreted as a
$3\sigma$ detection if $E_o > 3
\sqrt{\langle (\Delta   \hat{E})^2   \rangle}$.
The simulations show that a $3-\sigma$ detection is not possible 
for $R_b \leq 6 \,{\rm Mpc}$ and $R_b \leq 12 \,{\rm Mpc}$  
at  the GMRT and  MWA respectively. As noted earlier, the HI
fluctuations are somewhat under predicted in the simulations 
and the analytic predictions  $R_b \leq 8 \,{\rm Mpc}$  and $R_b \leq
16 \,{\rm Mpc}$ respectively, are somewhat larger. 

The limitation on the bubble size $R_b$ that can be detected 
is larger  for the MWA as compared to  the  GMRT. This is because of  
two reasons, the first being the fact that the MWA  has a very 
dense sampling of the small baselines where the HI fluctuation
are very large, and the second being the large FoV. 
In fact, the baseline distribution of the experiment has a significant
role  in determining the quantum of HI fluctuations and thereby
determining the lower cut-off for  bubble detection.
Looking for an optimum baseline distribution for bubble detection is
also an issue which we plan to address in future. In a situation where
the antenna layout is already in place, it may possible to tune the
filter to reduce the HI fluctuations. 

We have not considered the effect of peculiar velocities
\citep{bharad04} in our simulations.   From equation (22) of Paper I 
we see that the HI fluctuations scale as 
$\sqrt{\langle (\Delta \hat{E})^2   \rangle_{HF}} \propto
\sqrt{C_{l}}$, where $C_{l}$ is the  HI multi-frequency angular power   
spectrum (MAPS). The $C_l$s increase  by a  factor $\sim 2$ due to 
peculiar velocities, whereby $\sqrt{\langle (\Delta \hat{E})^2
  \rangle_{HF}}$ goes up by  a factor $\sim 1.5$. This  increase does
not significantly change our results, and is small compared to the
other uncertainties in the PR models.

The  signal  $\langle \hat{E} \rangle$ and the HI
fluctuations  $\sqrt{\langle (\Delta \hat{E})^2   \rangle_{HF}}$  both
scale as $\propto \bar{x}_{\rm HI}$, and the lower limit for bubble
detection  is unchanged for smaller neutral fractions.

\subsection{Size Determination}

In this subsection we estimate the accuracy to which it will be
possible to determine the bubble radius $R_b$. This, in general, is 
an unknown quantity that has to be determined from the
observation by trying  out filters with different values of 
 $R_f$. In the matched filter technique we  expect the predicted
SNR (only system noise) ratio   
\begin{equation}
{\rm SNR}=\frac{\langle\hat{E}\rangle}{\sqrt{\langle (\Delta \hat
    E)^2 \rangle_{\rm NS}}}
\label{eq:snr}
\end{equation}
to peak when the filter is exactly matched to the signal 
{\it ie.} $R_f=R_b$. The  solid line in the right panel of  Figure
\ref{fig:find_R10} shows this  for  $R_b=10\,{\rm
  Mpc}$. We find that the SNR peaks exactly when the  filter size 
$R_f=10\,{\rm   Mpc}$. We propose that this can be used to
observationally determine  $R_b$. 
For varying $R_f$, 
we consider  the ratio of the  observed value $E_o$ to the expected
system noise $\sqrt{\langle   (\Delta \hat     E)^2 \rangle_{\rm
    NS}}$, referring to this as  the  SNR. The $R_f$ value
where this SNR peaks gives  an estimate  of the actual bubble size $R_b$.  
The observed SNR will differ from the predictions of
eq. (\ref{eq:snr}) due to the  HI fluctuations outside the bubble.
These variations will differ from realization to realization and this 
can  introduce uncertainties in   size estimation.  We have used 
the simulations to estimate this.

The left panel of Figure \ref{fig:find_R10} shows the SNR as a
function of $R_f$ for $4$
different realizations of the simulation for the GMRT with bubble size 
$R_b=10 \, {\rm Mpc}$.  We see that for $R_f \leq R_b$ 
the SNR shows a very similar behavior in all the  realizations, and
it  always peaks  around $10 \,{\rm Mpc}$ as expected. For $R_f  > R_b$   
the  behavior of the SNR as a function of $R_f$ shows 
considerable variation  across the realizations. In some cases the
drop in SNR away from the peak is quite rapid whereas in others it is
very gradual (for example, the dashed-dot-dot curve). In many cases
there is an   spurious extra peak in the SNR at  an $R_f$ value that is
much larger than $R_b$. These spurious peaks do not pose a problem for
size determination as they are well separated from $R_b$ and can be
easily distinguished from the genuine peak. 

The error-bars in the right panel of   Figure \ref{fig:find_R10} show
the $3-\sigma$ fluctuation in the simulated  SNR determined from $24$
realizations of the simulation. Note that the fluctuations  at
different $R_f$  are correlated. Although the overall 
amplitude changes from one realization to another, 
the shape of the curve in the vicinity of $R_f=R_b$ is nearly
invariant across  all the realizations. In all of the $24$
realizations we can identify a well defined peak at the expected value 
$R_f=R_b$.  

Figures  \ref{fig:find_R20} and \ref{fig:find_R20M} show
the results for a similar analysis with $R_b=20 \, {\rm Mpc}$ for the
GMRT and the MWA respectively. It is not possible to detect a bubble
 of size $R_b=10 \, {\rm Mpc}$ with the MWA,  and hence we do not show
this. Here again,we find that  for all the realizations of the
simulations  the SNR peaks at $R_f=R_b$. The relative variations in
the SNR across the realizations is much less for $R_b=20 \, {\rm Mpc}$
as compared to $10 \, {\rm Mpc}$ and there are no spurious peaks. 
 Also, for the same bubble size  the
variations are smaller for the GMRT as compared to the MWA.  We do not
find any spurious peaks for $R_b=20 \, {\rm Mpc}$.

A point to note is that the mean SNR determined from the simulations
is somewhat smaller than the analytic predictions, both being shown in
the right panels of Figures \ref{fig:find_R10},  \ref{fig:find_R20} and
Figure \ref{fig:find_R20M}.  
 There are a couple of reasons that could account for this namely,
(i) the bubble in the simulation is  not exactly a sphere 
because of the finite grid size and  thus the match between the
filter and the signal is not perfect even when the sizes are same and
(ii)  the finite box-size imposes a minimum baseline beyond which the
signal is not represented in the simulation.

Based on our  results  we conclude that in the SB scenario for the
GMRT the accuracy to which the  bubble size can be determined  in our 
simulations is decided  by the resolution $2 \, {\rm Mpc}$ and not 
by the HI   fluctuations. In reality the limitation will come from the
angular resolution of the instrument which sets the limit at
$0.5 \, {\rm Mpc}$ for the GMRT and $8 \, {\rm Mpc}$ for the MWA. 

The height of the SNR peak  depends on the neutral fraction and it 
can be used to observationally determine this. We find that the HI
fluctuations do not change the position of the peak but introduce
considerable variations in its height even if $x_{\rm HI}=1$. 
The HI fluctuations  restrict the accuracy to which the 
neutral fraction can be estimated, an issue that we propose to address
in future work.

\subsection{Determining the position}

In the previous two subsections, we have considered cases where the
bubble's position  is known. Here we assume that the bubble's size is
known and we estimate the accuracy to which its position can be
determined in the presence of HI fluctuations. The situation
considered here is  a blind search whereas the former is a
targeted search centered on a QSO. 

In a real situations it would be necessary to jointly determine four 
parameters the bubble radius $R_b$, two angular coordinates
($\theta_x, \theta_y$) and the central frequency $\nu_c$ from the
observation. However, to keep the computational requirement under
control,  in this analysis we assume that $R_b$  is known. The
bubble is placed  at the center of the FoV   and frequency band, and
we  estimate how well  the position can be recovered  from the
simulation. To determine the bubble's position we move  the center of
the filter to different positions and search for a peak in the SNR. 
To reduce the computational requirement,   this is done along one
direction at a time, 
keeping the other two directions fixed at the bubble's actual center.   
We have also carried out simulations where the bubble is located 
off-center. We do not explicitly show these results because they are
exactly the same as when the bubble is at the center except for the
fact  that the value of the peak SNR is lower because of the
primary beam pattern.

Figure \ref{fig:find_theta_nu10} shows the results for $R_b=10\,
{\rm Mpc}$  for the GMRT. The left panel shows results for $4$  realizations of
the simulation, the right panels show the mean and $3-\sigma$  
determined from $24$ realizations of the simulations and the analytic
prediction for the mean value. In all cases a peak is seen at the
expected position matched with the bubble's actual center. 
The HI fluctuations pose a severe problem for determining the 
bubble's position as it introduces considerable fluctuations in the
SNR. In some cases these fluctuations are comparable to the peak at
the bubble's actual position     (see the dashed line in the upper
left panel). The possibility of  spurious peaks makes it difficult to
reliably determine the bubble's position. 
 
We present the results for  $R_b=20 \,{\rm Mpc}$
in Figures \ref{fig:find_theta_nu20} and  \ref{fig:find_theta_nu20M} 
for the GMRT and MWA respectively. The HI fluctuations do not pose a
problem for determining the position of such bubbles using the
GMRT. In all the realizations of the GMRT simulations there is a  peak
at the expected position. The FWHM  $\sim 40 {\rm Mpc}$ 
is approximately the same along $\theta_x$,$\theta_y$ and $\nu_f$  
and is comparable
to the separation at which the overlap between the bubble and the
filter falls to half the maximum value. The HI fluctuations does
introduce spurious peaks, but these are quite separated from the actual
peak and have a smaller height. We do not expect these to be of
concern for position  estimation.

The MWA simulations all show a peak at the expected bubble
position. The FWHM along $\theta $  ($\sim 60 \, {\rm Mpc}$ )
is somewhat  broader than that along $\nu$ ($\sim 40 \, {\rm
  Mpc}$).    The low spatial resolution $\sim 8 \, {\rm Mpc}$ possibly  
contributes to increase the FWHM along $\theta$. The HI fluctuations
introduce spurious peaks whose heights are  $\sim 50 \, \%$ of the
height of the actual peak.

\subsection{Bubble Detection in Patchy  Reionization }

The SB scenario considered till now is the most optimistic scenario in
which the HI traces the dark matter. The presence of 
ionized  patches other than the one that we are trying to detect
is expected to increase the contribution from HI fluctuations. 
We first consider the PR1 scenario where there are several additional
ionized bubbles of radius  $6 \, {\rm Mpc}$ in the FoV. 
Figures \ref{fig:R_Es_pr1} \& \ref{fig:R_Es_pr1M} show the mean  
value of the estimator  $\langle\hat{E}\rangle$ and $3-\sigma$
error-bars  as a function of $R_f$ for the GMRT and the MWA 
respectively. These were estimated from $24$ different realizations of
the simulation, using a  filter  exactly
matched to the bubble. 

We find that the results are very similar to those for the SB scenario
except that  the signal is  down by $0.6$ due to the lower neutral fraction
($x_{\rm HI}=0.62$) in the PR scenarios. Ionized bubbles with radius 
$R_b=8\, {\rm Mpc}$ and  $=12\,{\rm Mpc}$ or smaller cannot be detected
by the GMRT and MWA respectively due to the HI fluctuations. These
limits are similar to  those obtained in  simulations of  the SB
scenario.  

In the PR2 scenario the FoV contains  other  ionized bubbles of the
same size as the bubble that we are trying to detect.  We find that
bubble detection 
is not possible in such a situation, the HI fluctuations always
overwhelm the  signal.   This result obviously depends on 
number of other  bubbles in the FoV,  and this is decided by 
$x_{\rm HI}$ which we take to be $0.62$. A detection may be possible
at higher $x_{\rm HI}$ where there would be fewer  bubbles in the FoV.

In the SM scenario, the very large computation time restricts us from
generating several realizations  with central ionized regions of
different sizes. Hence we are unable to study the  restriction on
bubble detection.    We have only three realizations all of which have  
the same ionized region located at  the center of the box.  Based on
these we find that the  mean estimator $\langle
\hat{E} \rangle$ is $\sim30$ times larger that the standard deviation
due to HI fluctuations. The  detection of a bubble  of the size
present in our simulation (Figure \ref{fig:image}) is not
restricted by the HI fluctuations. We present size determination
results in  Figure \ref{fig:find_R42}. We see that the SNR peaks at
$R_f=42\, {\rm Mpc}$ and not at $R_f=27\, {\rm  Mpc}$. Recall  that
in the 21-cm map (Figure \ref{fig:image}) we have visually identified
the  former  as the bubble's  outer radius which includes
several small patchy  ionized regions towards the periphery 
 and the latter is the  inner radius which encloses  
the  completely   ionized region. We see that the matched  filter
identifies the  bubble's outer radius. 
To study the effect of non-sphericity we compare our results in  
Figure \ref{fig:find_R42}  with the 
predictions for  a spherical bubble of radius $R_b=42\, {\rm Mpc}$
embedded in uniform HI with  the same neutral fraction $x_{\rm
  HI}=0.5$. We find that our  results for the SM scenario follow the
spherical bubble prediction up to a filter size $R_f=28\, {\rm
  Mpc}$  (marked with a vertical line in
Figure \ref{fig:find_R42}), corresponding to the bubble's inner
radius which encloses a perfectly 
 spherical ionized region. Beyond this, and upto the outer
radius of  $42 \, {\rm Mpc}$, the HI is not fully ionized. There are 
 neutral patches  which introduce deviations from
spherical symmetry and cause the SNR to fall below   the
predictions of a spherical bubble beyond  $R_f=28 \, {\rm Mpc}$. The 
deviations from sphericity also broadens the  peak in the SNR
relative to the predictions for a spherical bubble. 

Our results based on the SM scenario show that the matched filter
technique works well for bubble detection and for determining the
bubble's size even when there are deviations from sphericity. We
obtain good estimates for the extents of  both,  the completely ionized
region and the partially ionized region.  For the SM scenario, 
 Figure \ref{fig:find_theta_nu42} shows   how well the bubble's 
position can be determined in a blind search. 
We have followed the  same  method as  described for the SB scenario
in  subsection 3.3. We  see that the SNR peaks at  the  expected
position. Further,  as  the bubble size is quite large $\gtrsim 
27$ Mpc there are no spurious peaks.

\section{Redshift Dependence}
Results shown so far are  all at  only one redshift $z=6$. It would
be  interesting and useful  to have  predictions for higher 
redshifts. However, addressing this issue through direct computations
at different redshifts would require considerable  computation 
beyond the scope of this work. Since we find that the analytic
predictions of Paper~I  are in good agreement with the simulations of
the SB scenario, we use the analytic formalism to predict how
different quantities are expected to scale with increasing $z$.

 The redshift dependence of some of the quantities 
like  the system noise, the background 21-cm brightness
$\bar{I_{\nu}}$,  and the angular and frequency extent of a bubble of
fixed comoving radius  causes the SNR to decrease with increasing
$z$. On the other hand the $z$ dependence of the 
neutral fraction, the  baseline distribution  function 
  and the effective antenna collecting area 
acts to increase the  SNR at higher redshifts.
 We find that with increasing $z$ both  $\langle \hat{E}\rangle$ and
 $\sqrt{\langle  (\Delta \hat E)^2 \rangle_{\rm  HF}}$  decrease  by
nearly  the same  factor  so that the   restriction on bubble
 detection does not change significantly at  higher redshift in the SB
 scenario. Assuming that the  neutral fraction does not change with
 $z$, the SNR for bubble detection decreases with increasing 
 redshift,  the change depending  on the bubble size. For  example, for
 the GMRT at $z=10$  the  SNR decreases by a factor $\sim 7$ and $6$
 for bubbles  of size   $R_b=10$ and $20\, {\rm Mpc} $
 respectively. For the  
 MWA this factor is $3$ for both these bubble sizes. 
We expect a similar  change in the SNR for the patchy reionization
scenarios.  The drop  in SNR is slower for the MWA relative to the GMRT 
because the effective antenna collecting area of the
 MWA increases at higher redshifts.   

The SNR is directly proportional to the global   neutral fraction $
x_{\rm HI}$  which increases  with $z$.  The details of how $x_{\rm
  HI}$ and the HI   fluctuations change with redshift depends on how 
reionization proceeds with  time, an issue beyond the scope of this
paper.

\section{Summary}
We have used a visibility-based formalism, introduced in 
\cite{datta2}, to simulate the detection of  spherical HII bubbles in
redshifted 
21 cm maps through a matched-filtering technique. The main aim
of this work is to use simulations to quantifying 
the limitations for bubble  detection  arising from the HI
fluctuations outside the bubble. We have computed 
the results for two instruments, namely, the  GMRT and the upcoming MWA.
Our main conclusions are as follows:

\begin{itemize}
\item In the case where the HI fluctuations outside the bubble
trace the dark matter distribution (SB scenario), 
we find that bubbles with radius 
$R_b=6\, {\rm Mpc}$ and  $=12\,{\rm Mpc}$ or smaller cannot be detected
by the GMRT and MWA respectively due to the HI fluctuations. Note that
this limitation is fundamental to the observations and cannot be
improved upon by increasing integration time.

\item For targeted observations of ionized bubbles, the bubble size
can be determined to an accuracy limited by the instrument's 
resolution; we find that HI fluctuations do not play any significant role.
However, the HI fluctuations can restrict the accuracy to which the 
neutral  fraction can be estimated. In addition, we find that
determining the position of the bubble in a blind search could be
quite difficult for small ($\sim 10 \, {\rm Mpc}$) bubbles as the HI
fluctuations introduce large fluctuations in the signal; for larger bubbles
the accuracy is determined by the instrument's resolution.

\item In a scenario of patchy reionization where the targeted HII
region is surrounded by many
small ionized regions of size $\sim 6 \, {\rm   Mpc}$ (PR1 scenario),
the lower  limit for bubble detection  is  similar to that in the SB
scenario.  
Thus the assumption that the HI traces the
dark matter gives a reasonable estimate of the contribution from HI
fluctuations if the background ionized bubbles are small 
 $\sim 6 \,{\rm Mpc}$. 
However, the situation is quite different when the surrounding bubbles  
as of similar size  as the targeted bubble (PR2 scenario). The large
HI fluctuations do not permit bubble detection for a neutral fraction
$x_{\rm HI}<0.6$.  Thus  for $x_{\rm HI}=0.6$  or lower,  bubble
detection is possible only if the other  ionized regions in the 
FoV  are much smaller than the bubble that we 
are trying to detect.

\item  The matched filter technique works well for 
more realistic cases based on the semi-numeric modelling of 
ionized regions \citep{choudhury08}. Here  the bubbles are substantially
non-spherical because of surrounding bubbles and inhomogeneous
recombination. Our method  gives a good estimate of the size of both
the  fully ionized and the partially ionized regions in the bubble. 

\begin{figure}
\includegraphics[width=.3\textwidth, angle=270]{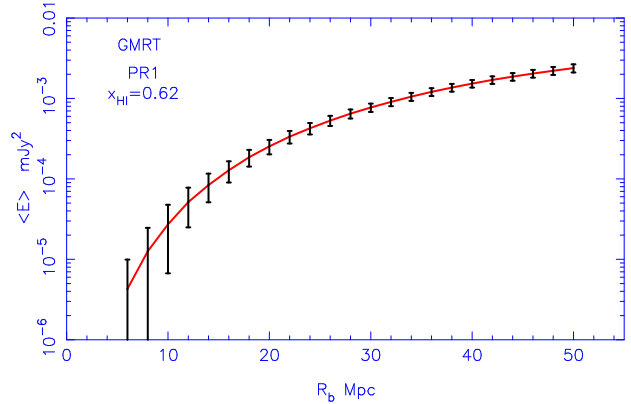}
\caption{The mean    $\langle\hat{E}\rangle$ and $3-\sigma$
  error-bars   of the estimator as a function of   $R_f$ for the GMRT
 estimated from    $24$ different realizations of the   PR1 scenario. 
In all cases the filter  is   exactly matched to the bubble.}
\label{fig:R_Es_pr1}
\end{figure}

\end{itemize}

To put our conclusions in an overall perspective, let us consider 
an ionized bubble around a luminous QSO at $z \gtrsim 6$. 
We expect $R_b \gtrsim 30 {\rm Mpc}$ 
from studies of QSO  absorption spectra (\citealt{wyithe04b,mesinger04}). 
It has also
been pointed out that these bubbles may survive as large ``gray fossils''
a long time after the source has shut down (\citealt{furlanetto08}).
It will be possible  to detect such bubbles only if 
the background bubbles are smaller, say, 
$< 30 \, {\rm Mpc}$. We find from  
models of \cite{mesinger07} 
that the typical sizes of ionized regions
when $x_{\rm HI} \sim 0.3 (0.1)$ is $\sim 20 (70) {\rm Mpc}$. Though 
these values could be highly model-dependent, it still gives 
us an idea that the bubbles around the  luminous
QSO would be detectable even in a highly ionized
IGM with, say, $x_{\rm HI} \sim 0.3$. If the size of the targeted
bubble is larger, then this constraint is less severe. This gives
a realistic hope of detecting these bubbles at $z \gtrsim 6$ with 
near-future facilities.

A caveat underlying a large part  of our analysis is  the
assumption that the bubbles  under consideration are perfectly
spherical. This is note the case in reality. For example,
non-isotropic emission from the sources (QSOs), density fluctuations
in the IGM and  radiative transfer effects would distort the 
shape of the bubble. The semi-numeric simulations (SM scenario)
incorporate some of these effects and give an  estimate  of the impact of
the deviations from sphericity on bubble detection.  This is
an important issue which we plan to address in more detail in future. 
In addition, the   finite light travel time gives rise to an 
apparent non-sphericity even if the physical shape is spherical
\citep{wyithe04a,yu05}.  
This effect can, in principle, be estimated analytically  and
incorporated in  the filter. We plan to address this effect
in a separate publication.

\begin{figure}
\includegraphics[width=.3\textwidth, angle=270]{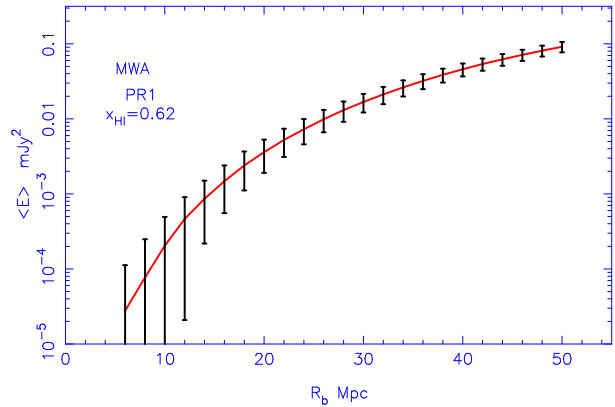}
\caption{Same as the Figure \ref{fig:R_Es_pr1} for the MWA.}
\label{fig:R_Es_pr1M}
\end{figure}

\begin{figure}
\includegraphics[width=0.4\textwidth, angle=270]{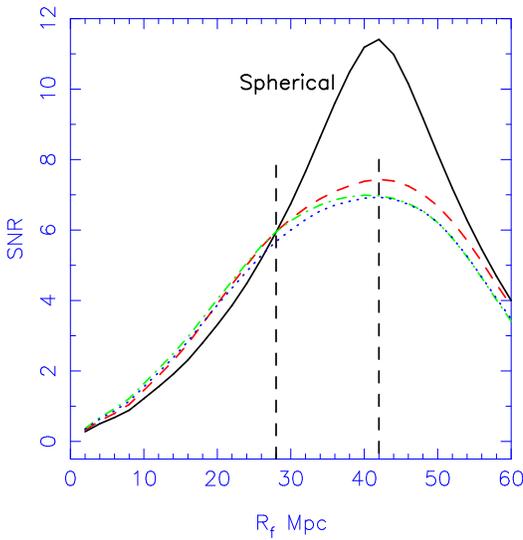}
\caption{Same as  Figure \ref{fig:find_R10} for the SM scenario for
  the GMRT.  The dotted, dashed dotted and dashed lines show
  results for three different realizations.  To show the 
  effect of non-sphericity, we compare these results
  with predictions for a  spherical bubble of sized $R_b==42\, {\rm
    Mpc}$ embedded in  uniform HI   with neutral fraction $x_{\rm
  HI}=0.5$(solid line).  The vertical line at $R_f=28\, {\rm 
    Mpc}$ shows the radius up to which the bubble is fully ionized and
  the SNR   follows the spherical predictions. The SNR peaks at
  $R_f=42\, {\rm     Mpc}$ marked by another vertical line.}  
\label{fig:find_R42}
\end{figure}

\begin{figure*}
\includegraphics[width=.35\textwidth, angle=270]{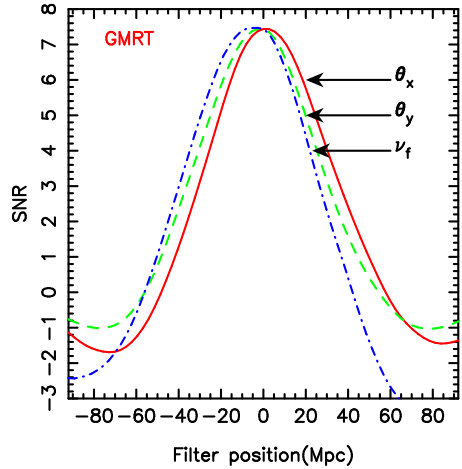}
\caption{Same as  Figure \ref{fig:find_theta_nu10} for the SM
  scenario for the GMRT. The x-axis shows the comoving distance of
  the filter position from the center of the box. The three curves 
respectively show  results for a  search along three $\theta_x$,
  $\theta_y$ and  $\nu$  axes.} 
\label{fig:find_theta_nu42}
\end{figure*}

\section{Acknowledgment}
We would like to thank M. Haehnelt and J. Regan for allowing us
to use the ionization maps of \cite{choudhury08}.
KKD and SM would like to thank  Prasun Dutta and Prakash Sarkar for
useful discussions.  KKD is supported by a senior research fellowship
of Council of  Scientific and Industrial Research (CSIR), India.

\end{document}